\documentclass[aps,pra,groupedaddress,showpacs,reprint]{revtex4-1}
\usepackage{graphicx}
\usepackage{bm}

\begin{document}

\title{Potential splitting approach to multichannel Coulomb scattering: the driven Schr\"odinger equation formulation }

\author{M.~V.~Volkov$^{1,2}$}
\email[E-mail: ]{miha@fysik.su.se}
\author{S.~L.~Yakovlev$^3$}
\email[E-mail: ]{yakovlev@cph10.phys.spbu.ru}
\author{E.~A.~Yarevsky$^3$}
\email[E-mail: ]{yarevsky@gmail.com}
\author{N.~Elander$^1$}
\email[E-mail: ]{elander@fysik.su.se}

\affiliation{$^1$ Department of Physics, AlbaNova University Center, Stockholm University,
106 91 Stockholm, Sweden}

\affiliation{$^2$ Department of Quantum Mechanics, St Petersburg State University, 198504 St Petersburg, Russia}

\affiliation{$^3$ Department of Computational
Physics, St Petersburg State University, 198504 St Petersburg, Russia }

\begin{abstract}

{In this paper we suggest a} new approach for the multichannel Coulomb scattering problem.
The Schr\"{o}dinger equation for the problem is reformulated in the form of a set of inhomogeneous equations with
{a finite-range driving term}.
The boundary {conditions at infinity for this} set of equations have been proven to {be purely} outgoing waves.
The formulation {presented here} is based on splitting the interaction potential {into} a finite range core part
{and a long} range tail part. The conventional matching procedure coupled with the integral Lippmann-Schwinger equations
technique {are used in the formal theoretical basis of this approach. The reformulated scattering problem is suitable for
application in the exterior complex scaling technique: the practical advantage is that after the complex scaling the problem is
reduced to a boundary problem with zero boundary conditions. The Coulomb wave functions are used only at a single point: if this
point is chosen to be at a sufficiently large distance, on using the asymptotic expansion of Coulomb functions, one may completely
avoid the Coulomb functions in the calculations.
The theoretical results are illustrated with numerical calculations for two models.}

\end{abstract}


\maketitle

\section{Introduction}
The three-body Coulomb scattering problem remains {both} a formally as well as computationally challenging problem  \cite{R:Rshakeshaft09}.
{An} understanding of three-body scattering has {implications to many fields of science, for examples, to} combustion studies
\cite{appcombust}, reactions in the interstellar medium \cite{appinststel}, plasma chemistry \cite{applasma} and other types of gas phase
reactions
\cite{apgasphas}. Besides {their} fundamental interest, these applications have {also} generated
{significant experimental efforts}. {The double electrostatic storage ring (DESIREE) currently
being constructed at Stockholm university \cite{r:desiree} is designed to investigate the gas-phase reactions
of oppositely charged species, of which three-body scattering reactions are the experimentally simplest subset,
so generating high-quality data which will challenge theoretical studies as well as providing much-needed input
to models of interstellar and plasma reactions.}

In Ref.~\cite{VEYY09} {we initiated a set of studies with the aim of obtaining} a method for accurately
computing state selective three-body multi channel scattering which {also included} Coulomb interaction.
Our method was inspired by {the} mathematically sound approach of Nuttall and Cohen~\cite{NutCoh} {related to}
exponentially decreasing or finite range potentials. Rescigno {\it et al.} \cite{RBBC97} modified this approach to scattering
by non-Coulombic but long-range potentials. The recent formalism developed in \cite{VEYY09} provides the mathematically solid
basis for {application of the complex} scaling method \cite{NutCoh, RBBC97} to the single channel Coulomb scattering problem.

In Ref.~\cite{EVLSMYY09} we outline how this new {formalism can be extended to problems in three-body scattering which
include Coulomb interactions}. This {extension is based} on the same formal and numerical technique as {was}
used for computing three-body resonances \cite{Alferova}.  {Furthermore, this} extension {can be also combined}
with our three-body resonance methods \cite{Alferova, ELY2001} into a technique by which one may quantitatively identify the
influence of resonances \cite{ksh_nh3} {on} the three-body scattering cross section. The purpose of the present
contribution is to {study in detail the} advantages and {limitations} of the present formulation of two-body
multi channel Coulomb scattering. We are {currently investigating the possibility to extend this formulation to one that}
can be generalized to accurate studies of three-body multi channel scattering where collisions between charged fragments
{occur} \cite{r:desiree}.

{Formally, the new} formalism is based on splitting the entire potential in a sum of two sharply cut-off
{potentials: a core potential, which is then of finite range; and} a tail potential, where all but the Coulomb
interaction can be neglected. In a recent contribution \cite{Yakovlevetal} we have studied the structure of the solutions
of the three-dimensional Schr\"{o}dinger equation for sharply cut-off Coulomb potentials. The derived formulation of
the three-dimensional driven Schr\"{o}dinger equation for problems involving Coulomb interaction is shown to open the way
for the forthcoming applications in three-body systems as outlined in Ref.~\cite{EVLSMYY09}.

The report {is structured as follows}. The main ideas behind this work are described in section~\ref{inttheor}.
We begin by presenting the two-body multichannel scattering problem in subsection~\ref{2bScPr}.
The analytically solvable equations for the diagonal part of the potential tail are investigated in subsection~\ref{ansolv} while the formulation of the problem for the total potential tail is found in subsection~\ref{pottail}.
The complex scaling theory is combined with the scattering theory in subsection~\ref{DSEECS}.
The numerical implementation of our theory is presented in section~\ref{Num}.
Two different examples are considered in subsections~\ref{Noro} and~\ref{HN}.
{Finally, section~\ref{sum} presents a summary of the report.}

\section{Theory}
\subsection{Theoretical background}\label{inttheor}
The boundary conditions for a scattering problem are conventionally specified as {the} superposition of
an incident wave and an outgoing wave. For two-body systems with central interactions the Schr\"odinger equation
can be reduced to a set of one-dimensional partial-wave equations. The boundary conditions for a single one-dimensional
equation can easily be  specified and numerically implemented. For the three body scattering problem the boundary conditions
at infinite distances between {the} particles are much more complicated. As such it is difficult to implement these
boundary conditions in practical calculations.

{One of the techniques commonly used to avoid the problems arising from these boundary conditions is the complex
scaling transformation~\cite{BC71} method.} In this method one maps the radial coordinate $r$ onto a path $g_{\alpha}(r)$ in the
upper half of the complex coordinate plane
\begin{eqnarray}
\label{CS}
&&r \longmapsto g_{\alpha}(r)~, \qquad 0< \alpha \le \pi/2~,
\nonumber\\
&&g_{\alpha}(r) \sim \mbox{const} + r\exp(i\alpha)~, \qquad\mbox{as}\qquad r\to \infty~.
\end{eqnarray}
This method is widely and successfully used when calculating resonance energies {in problems} where the boundary
condition at infinity is a purely outgoing wave. It is well known that the transformation (\ref{CS}) converts a purely
outgoing wave to an exponentially decreasing function. Thus, if  the scattering problem is reformulated in such a way
that the boundary condition at infinity is a purely outgoing wave then, after a complex scaling {transformation, one}
obtains a problem with zero boundary {conditions} at infinity.

This approach to {scattering problems was first reported} by Nuttall and Cohen in 1969~\cite{NutCoh}.
{These authors suggest applying the Hamiltonian operator to the difference between the scattering solution of the Schr\"{o}dinger
equation and an incident wave to yield an inhomogeneous (driven) Schr\"{o}dinger equation.} This {difference between the}
functions apparently behaves as a purely outgoing wave at infinity. Hence, this equation is suitable for {complex scaling,
and \cite{NutCoh} used the uniform complex scaling approach}
\begin{eqnarray}
\label{UCS}
g_{\alpha}(r) = r\exp(i\alpha)~.
\end{eqnarray}
{However, it 
appears that this approach is limited to the cases of finite-range and exponentially decreasing potentials.}
This is due to the following reason. The inhomogeneous (driving) part of the driven Schr\"{o}dinger equation is the product of the
potential energy and the incident wave. The incident wave after the complex scaling transformation becomes a superposition of the
increasing and decreasing exponential functions at infinity. Hence, the entire inhomogeneous part diverges. The class of finite-range
and exponentially decreasing potentials form an exception where the method of Ref.~\cite{NutCoh} can be successfully applied.

The formulation of Nuttall and Cohen was modified by Rescigno {\it et al} \cite{RBBC97} for one-dimensional single-channel long-range
potentials ({explicitly except Coulomb potentials}). These authors also start from the equation for the difference between
{the} scattering solution of the Schr\"{o}dinger equation and an incident wave. However, instead of the potential $V(r)$ they
use the finite-range potential $V_R(r)$ defined as
\begin{eqnarray}
\label{V_R}
V_R(r) =
\left\{
\begin{array}{lll}
V(r)&,&r \le R~,\\
0&,&r>R~.
\end{array}
\right.
\end{eqnarray}
The driven Schr\"{o}dinger equation with this potential does not experience difficulties with divergence after {applying} complex
scaling since this potential is of finite-range. {Furthermore}, the solution of the unscaled problem with the truncated potential
$V_R(r)$ approaches the solution of the original problem with the entire potential $V(r)$ as $R\to\infty$. As shown in~\cite{RBBC97}
the same is not true for the scaled equation. The solution of the scaled equation with the potential $V_R(r)$ gives the incorrect
scattering amplitude since the function (\ref{V_R}) is not analytic \cite{BC71}. Therefore, the authors of \cite{RBBC97} suggested
to use exterior complex scaling~\cite{HisSig,S79}. This transformation belongs to the more general type given by Eq.~(\ref{CS}).
It is defined by the function $g_{\alpha,Q}$ where $Q$ is the scaling point. The inner interval $[0,Q]$ is mapped {onto} itself
\begin{eqnarray}
\label{ECS}
g_{\alpha,Q}(r) = r~, \qquad r \le Q~.
\end{eqnarray}
If $Q \ge R$, the solution of the scaled equation with the truncated potential $V_R(r)$ approaches the solution of the scaled equation
with the entire potential $V(r)$ as $R\to\infty$. One can calculate the scattering amplitude with desired accuracy {by} choosing
a proper value of $R$.

This method is not applicable to the Coulomb scattering problem directly since as it is well known {that} truncation of the Coulomb
potential leads to noticeable errors for any truncation radius $R$. In the two body scattering problem the Coulomb potential can be implemented
into the discussed approach if it is included in the free-motion Hamiltonian, while $V(r)$ describes the short-range part of the interaction.
In this case the incident wave is represented by a Coulomb wave function, which is known analytically. This approach has been successfully
used for calculations in atomic~\cite{McCurdy2004} and nuclear~\cite{Kruppa} physics. Unlike the two body case, {an} analytic solution
for the Coulomb problem does not exist if three or more particles are involved in the scattering process.

In our recent report~\cite{VEYY09} we have shown how the method of exterior complex scaling can be generalized {to} the Coulomb
scattering problem. Instead of truncating the potential we {represent} the entire potential as $V(r) = V_R(r) + V^R(r)$, where
$V_R(r)$ is the same as in Eq.~(\ref{V_R}) and the potential tail  $V^R(r)$ is given by
\begin{eqnarray}
\label{V^R}
V^R(r) =
\left\{
\begin{array}{lll}
0&,&r \le R~,\\
V(r)&,&r>R~.
\end{array}
\right.
\end{eqnarray}
{The approach discussed in this recent report~\cite{VEYY09}} is based on solving the problem for the potential tail $V^R(r)$
at the first {step}. The solution of the scattering problem for the potential tail $V^R(r)$ plays the role of the incident wave.
{On subtracting} this incident wave from the scattering wave function we {obtain} a function which asymptotically behaves
as a purely outgoing wave. After the transformation (\ref{CS}) with the function $g_{\alpha,Q}$ satisfying {(\ref{ECS}),}
the boundary problem for this function has the trivial zero boundary conditions both at the origin and at infinity.

In the present study we proceed with a formal as well as numerical study of the two-body single and multichannel problems.
{All the potentials discussed in this treatment that are denoted with a subscript or superscript $R$ are defined analogously
to $V_R$ and $V^R$ in Eqs. (\ref{V_R}) and (\ref{V^R}), respectively.}

\subsection{The two-body multichannel scattering problem}\label{2bScPr}

{In the following discussion, consider the} two-particle multichannel scattering problem with $M$ channels.
We assume that the interaction between the particles ${\cal V}_{nm}$ ($n,m = 1,\ldots,M$) depends only on the {inter-particle} distance
{$r$ and when $r\to\infty$ can be asymptotically represented as}
\begin{eqnarray}
\label{thresh}
{\cal V}_{nm}(\infty) = \delta_{nm}t_n~. 
\end{eqnarray}
The quantities $t_n = {\cal V}_{nn}(\infty)$ are called the thresholds. The total interaction is given by {the} sum
\begin{eqnarray}
\label{totpot}
{\cal V}_{nm}(r) = \delta_{nm}\frac{Z_1^nZ_2^n}{r} + {\cal V}^s_{nm}(r)+\delta_{nm}t_n~.
\end{eqnarray}
The first diagonal term corresponds to the Coulomb interaction while ${\cal V}^s_{nm}(r)$ describes the short-range
interaction which is assumed to decrease faster than $r^{-2}$ {for large particle separations}. More precisely ${\cal V}^s_{nm}(r)$
should obey the condition
\begin{equation}
\int_0^\infty dr\, (1+r)|{\cal V}^s_{nm}(r)| < \infty .
\label{srpotV}
\end{equation}
The partial wave multichannel Schr\"{o}dinger equation for a given angular momentum $\ell$ (see for example~\cite{Newton,GoldWat,Taylor})
has the form of a set of equations for partial wave functions $\Psi_{fi}^{\ell}(r)$

\begin{eqnarray}
\label{SE2}
\left[-\frac{d^2}{dr^2}+\frac{\ell(\ell+1)}{r^2} + \frac{2k_f\eta_f}{r} - k_f^2\right]\Psi_{fi}^{\ell}(r)
\nonumber\\
+ \sum\limits_{n=1}^M V_{fn}(r)\Psi_{ni}^{\ell}(r)  =0~.
\end{eqnarray}
%
%
Here the coupling terms are of the form  $V_{nm}(r)= (2\mu/\hbar^2) {\cal V}^s_{nm}(r)$,
{while} the momentum  $k_n$ and the Sommerfeld parameter $\eta_n$ in the $n$-th channel are defined through the energy $E$
{by the expressions} $k^2_n=2\mu (E-t_n)/\hbar^2$ and $\eta_{n} =  Z^n_1Z^n_2\,\mu/(k_{n}\hbar^2)$.

The partial wave functions $\Psi_{fi}^{\ell}(r)$  satisfy the regularity condition at the origin
\begin{eqnarray}
\label{BCzero}
\Psi_{fi}^{\ell}(0) = 0~,
\end{eqnarray}
while, as $r\rightarrow\infty$, they have the asymptotics 
\begin{eqnarray}
\label{BC8}
\Psi_{fi}^{\ell}(r)  \sim \delta_{fi}e^{i\sigma^i_{\ell}}F_{\ell}(\eta_i,k_i\, r)+ {u^+_{\ell}(\eta_f,k_{f}\, r)A^s_{fi}}~.
\end{eqnarray}
Here the functions
$$
u^{\pm}_{\ell}(\eta_{n},k_{n}r) = e^{\mp i\sigma^n_{\ell}}\left[G_{\ell}(\eta_{n},k_{n}r)\pm iF_{\ell}(\eta_{n},k_{n}r)\right]
$$
are defined by using the regular (irregular) Coulomb wave {function $F_{\ell}\,(G_{\ell})$,
and ${\sigma^n_{\ell}} = \arg\left\{\Gamma(1+\ell+i\eta_{n})\right\}$ is the Coulomb phase shift
in the $n$-th channel~\cite{AbrSt}. It should be noted that if  the Coulomb interaction is not {present}
in a channel then the Coulomb  wave functions
in the {asymptotics}  (\ref{BC8}) 
should be replaced by the
Riccati-Bessel $ {\hat j}_{\ell}$ and Riccati-Hankel ${\hat h}^{\pm}_{\ell}$ functions \cite{AbrSt}
since  $F_{\ell}(0,k_{n}r) = {\hat j}_{\ell}(k_{n}r)$
and $u^{\pm}_{\ell}(0,k_{n}r) = {\hat h}^{\pm}_{\ell}(k_{n}r)$.

The {quantities} $A_{fi}^s$ are the scattering amplitudes due to the short-range interaction $V$.
Their dependence on $\ell$ and $E$ is assumed implicitly.
The total scattering amplitudes are given by the sum
$$
A_{fi} = A_{f}^{C}\delta_{fi}+A_{fi}^s
$$
where
\begin{equation}
A_{f}^{C}=\frac{\exp(2i\sigma^f_{\ell})-1}{2i}
\label{CoulombAmp}
\end{equation}
is the partial Coulomb scattering amplitude. The partial wave cross sections {are} then determined by the amplitude
through the standard expression
\begin{eqnarray}
\label{PCS}
{\sigma}^{\ell}_{fi} & = & \frac{k_f}{k_i^3}4\pi(2\ell+1){\left|{A_{fi}}\right|}^{\,2}~.
\end{eqnarray}
The total cross section corresponding to the reactive scattering transition $i\to f$ is given by the sum over momenta
\begin{eqnarray}
\label{TCS}
{\sigma}^{tot}_{fi} & = & \sum\limits_{\ell=0}^{\infty}{\sigma}^{\ell}_{fi}~.
\end{eqnarray}

{In the following discussion matrix notation is used. Here, the set of wave functions $\Psi_{fi}^{\ell}(r)$ are considered
as a square matrix with the indices $f$ and $i$ running over all values from $1$ to $M$. All matrices are denoted by bold typeface.
In this matrix notation,  Eq.~(\ref{SE2}) takes the form}
\begin{eqnarray}
\label{MSE}
\left[
{\bf H}_0 + {\bf L}(r) + {\bf C}(r) - {\bm k}^2\right]&&{\bf \Psi}(r) \nonumber\\
+ {\bf V}(r)&&{\bf \Psi}(r) =0~.
\end{eqnarray}
Here  ${\bm k}$ and ${\bf H}_0$ represent  the diagonal matrices $k_n\,\delta_{nm} $ and $-d^2/dr^2  \delta_{nm}$ {, respectively}.
The centrifugal term  ${\bf L}(r)$, the Coulomb interaction {term} ${\bf C}(r)$ and the short range potential coupling matrix
${\bf V}(r)$ are defined by the following matrix elements  $ \ell(\ell+1)/r^2\, \delta_{nm}$,  $2k_n\eta_n/r\, \delta_{nm} $ and $V_{nm}(r)$,
respectively. The total Hamiltonian matrix {is given by}
$$
{\bf H} =  {\bf H}_{0} + {\bf L}(r) + {\bf C}(r) + {\bf V}(r).
$$
The regularity condition (\ref{BCzero}) takes the matrix form
\begin{eqnarray}
\label{MBCzero}
{\bf \Psi}(0) = 0~.
\end{eqnarray}
By introducing the diagonal matrix $ [{\bm u}(r)]_{nm} =  u_{\ell}^{+}(\eta_n,k_n \,r)\delta_{nm}$ the boundary condition (\ref{BC8})
can be rewritten as follows
\begin{eqnarray}
\label{MBC8}
{\bf \Psi}(r)  \sim {\bf F}(r) + {\bm u}(r){\bm A}^s, \qquad r\rightarrow\infty~,
\end{eqnarray}
where ${\bf F}(r) = [{\bm u}(r){\bf D}-{\bm u}^*(r)]/(2i)$ and  $ [{\bf D}]_{nm} = \exp[2i\sigma^n_{\ell}]\delta_{nm}$.
{Due to the short-range interaction, ${\bf V}$, the amplitude matrix ${\bm A}^s$ is} constructed from the partial-wave
scattering amplitudes $A_{fi}^s$ with all possible values of indices $f$ and $i$. The total scattering amplitude matrix is then given by the sum
\begin{equation}
{\bm A} = {\bm A}^C + {\bm A}^s ,
\label{totampl}
\end{equation}
where the Coulomb scattering amplitude matrix (\ref{CoulombAmp}) is denoted as ${\bm A}^C$.

Solving Eq. (\ref{MSE}-\ref{MBC8}) by the exterior complex scaling {technique requires a} reformulation of the problem.
{In a similar approach to that employed in the one channel case}
\cite{VEYY09, Yakovlevetal} the new incident wave which incorporates the long range tail of the interaction potential should be
constructed on the first step. This approach will be described in the next two subsections.


\subsection{\textbf{The solution to the problem of the long range diagonal part of the potential}
}\label{ansolv}

Here we consider the diagonal part of the equations (\ref{MSE}) {with the interactions due to the long-range tail}
\begin{equation}
\left[{\bf H}_0 + {\bf L}(r) +{\bf C}^R(r) - {\bm k}^2\right]{\bm \psi}^{R}(r)=0
\label{MSElong}
\end{equation}
with the regularity boundary condition
\begin{equation}
{\bm \psi}^R(0)=0
\label{bcz}
\end{equation}
and with {the asymptotics } as $r\rightarrow\infty$
\begin{equation}
\label{bcas}
{\bm \psi}^R(r)  \sim {\bf F}(r) + {\bm u}(r){\bm A}^R.
\end{equation}
Solving the problem (\ref{MSElong}-\ref{bcas}) is naturally reduced to the construction of the solutions to the individual equations
\begin{equation}
\left[-\frac{d^2}{dr^2}+\frac{l(l+1)}{r^2}+ \frac{2k_{n}\eta_{n}}{r}\,\theta(r-R)-k_n^2\right]\psi^{R}_n(r)=0,
\label{diag SE}
\end{equation}
where $\theta(t)$ is the Heaviside step function defined such that $\theta(t)=0,\ t\le 0$ and $\theta(t)=1, \ t>0$. The diagonal
Coulomb tail potential $C^R_{nn}(r) $ {is} represented in Eq. (\ref{diag SE}) by its explicit form $2k_{n}\eta_{n}/r\,\theta(r-R)$.
The scattering solution to  Eq. (\ref{diag SE}) can be constructed {using the matching procedure that is described in detail}
in \cite{VEYY09, Yakovlevetal}. If $r\le R$ then $\psi^{R}_n(r)$ takes the form
\begin{equation}
{\psi}^R_n(r)={\hat j}_\ell(k_n r)a^R_n,
\label{psileR}
\end{equation}
{or in matrix notation}
\begin{equation}
{\bm \psi}^R(r)={\bf {\hat {j}}}(r){\bm a}^R
\label{psileRm}
\end{equation}
and if $r> R$ then $\psi^{R}_n(r)$ is given by
\begin{equation}
{\psi}^R_n(r)= e^{i\sigma^n_{\ell}}F_{\ell}(\eta_n,k_n r)+u^{+}_{\ell}(\eta_n,k_n r)A^R_n,
\label{psigrR}
\end{equation}
{or in matrix notation}
\begin{equation}
{\bm \psi}^R(r)= {\bf F}(r)+{\bm u}(r){\bm A}^R.
\label{psigrRm}
\end{equation}

{The matrices} ${\bm a}^R$ and ${\bm A}^R$ in (\ref{psileRm}, \ref{psigrRm}) are diagonal.
The values of diagonal elements $a^R_n$ and $A^R_n$
follow from the matching conditions at the point $r=R$ and read
\begin{eqnarray}
\label{a012}
{a^R_n} &=&\frac{k_n}{{\cal W}_R\left(u^+_{\ell},{\hat j}_{\ell}\right)}~,
\end{eqnarray}
\begin{eqnarray}\label{A012}
A^R_n &=&\frac{\exp\left[2i\arg(a^R_n)\right]-\exp[2i\sigma^n_{\ell}]}{2i} ~,
\end{eqnarray}
where ${\cal W}_R(f,g)$ denotes the Wronskian $f(r)g'(r)-f'(r)g(r)$ calculated at $r=R$ for the functions $f=u^{+}_{\ell}(\eta_n,k_n r)$
and $g={\hat j}_\ell(k_n r)$. Finally, the diagonal matrix ${\bm \psi}^R(r)$ is defined by
\begin{eqnarray}
[{\bm \psi}^R(r)]_{nm}=\psi^R_{n}(r)\delta_{nm}.
\label{psi^R}
\end{eqnarray}

Another kind of the solution to the equation (\ref{MSElong})  is defined by the asymptotic condition \cite{VEYY09} as $r\to \infty$
\begin{equation}
{\bm u}^R(r)\sim {\bm u}(r).
\label{u^Ras}
\end{equation}
This solution can be constructed by the same matching procedure {just employed and this} results in the following form of the components
of ${\bm u}^R(r)$
\begin{equation}
u^R_{n}(r)= u^+_\ell(\eta_n,k_n r)
\label{u^R}
\end{equation}
for $r>R$ and
\begin{equation}
u^R_n(r)= {h}^{-}_{\ell}(k_n r){c}^R_n +{h}^{+}_{\ell}(k_n r){d}^R_n
\label{u^RleftR}
\end{equation}
for $r\le R$.
The coefficients $c^R_n$ and $d^R_n$ are given by
\begin{eqnarray}
c^R_n&=&{\cal W}_R(u_\ell^+,h_\ell^+)/{\cal W}_R(h_\ell^-,h_\ell^+)
\nonumber \\
d^R_n&=&{\cal W}_R(u_\ell^+,h_\ell^-)/{\cal W}_R(h_\ell^+,h_\ell^-),
\label{c^Rd^R}
\end{eqnarray}
where the Wronskians are computed for {the} functions $u^+_{\ell}(\eta_n,k_n r)$ and $h^{\pm}_{\ell}(k_n r)$ at the point $r=R$.
The diagonal matrix
\begin{equation}
[{\bm u}^R(r)]_{nm}=u^R_n(r)\delta_{nm}
\label{bold u^R }
\end{equation}
provides the solution to the Eq. (\ref{MSElong}).

The solutions ${\bm \psi}^R$ and ${\bm u}^R$ allow us to construct the Green's function ${\bm g}^R$ by the standard formula
\begin{equation}
{\bm g}^R(r,r')={\bm k}^{-1}{\bm \psi}^R(r_{<}) {\bm u}^R(r_{>}),
\label{G^R}
\end{equation}
where $r_>(r_<) = \mbox{max}(\mbox{min})[r,r']$. This is possible since the diagonal matrices ${\bm \psi}^R$ and ${\bm u}^R$ commute.
By construction this function obeys the equation
\begin{eqnarray}
\left[{\bf H}_0 + {\bf L}(r) +{\bf C}^R(r) - {\bf k}^2\right]{\bm g}^{R}(r,r')\nonumber \\
={\bf I}\delta(r-r'),
\label{greenMSElong}
\end{eqnarray}
where ${\bf I}$ denotes the {unit} matrix.

\subsection{The scattering problem for the entire potential tail}\label{pottail}

Let us now consider the scattering problem for the entire tail potential matrix ${\bf C}^R(r)+{\bf V}^R(r)$. The coupled Schr\"odinger
equation in this case reads
\begin{eqnarray}
\left[{\bf H}_0 + {\bf L}(r) + {\bf C}^R(r)- {\bm k}^2\right]&&{\bm \Psi}^R(r)=\nonumber \\
-{\bm V}^R(r)&&{\bm \Psi}^R(r).
\label{MSE^R}
\end{eqnarray}
Regularity at $r=0$ and asymptotic {as $r\to \infty$, the} boundary conditions take the form
\begin{eqnarray}
\label{MBCzero+-}
&&{\bf \Psi}^{R}(0) = 0~,\\
\label{MBC8+}
&&{\bf \Psi}^{R}(r) \sim {\bf F}(r)+ {\bm u}(r){\bf {\cal A}}^{R}.
\end{eqnarray}
The solution to Eqs. (\ref{MSE^R}-\ref{MBC8+}) {are conveniently obtained from the solution} of the Lippmann-Schwinger integral equation
\begin{eqnarray}
{\bf \Psi}^R(r)&=&{\bm \psi}^R(r)- \int\limits_R^\infty \, dr' {\bm g}^R(r,r'){\bf V}^R(r'){\bf \Psi}^R(r').
\label{LSE^R}
\end{eqnarray}
This equation uses {both the solution ${\bm \psi}^R$ and the Green's function ${\bm g}^R$ that are constructed
in the preceding subsection.}
Due to (\ref{srpotV}) {these type of equations have a} unique solution \cite{DeAlfaro}. From this equation it follows that
{${\bf \Psi}^R(r)$, similar to $ {\bm \psi}^R(r)$,} takes a different functional form if $r\le R$ or $r> R$.

Using (\ref{psileR}) for $r\le R$ {and the relevant representation of the Green's function} ${\bm g}^R(r,r')$ one arrives at the expression
\begin{equation}
{\bf \Psi}^R(r)={\bf {\hat j}}(r){\bf a}^R ,
\label{Psi^RleR}
\end{equation}
where the matrix ${\bf a}^R$ has the form
\begin{equation}
{\bf a}^R={\bm a}^R\left[{\bf I} - {\bm k}^{-1}\int\limits_R^\infty  d r'\, {\bm u}(r'){\bf V}(r'){\bf \Psi}^R(r')\right] .
\label{aa^R}
\end{equation}

For $r>R$, the equation (\ref{LSE^R}) reads
\begin{eqnarray}
{\bf \Psi}^R(r)&=&{\bm \psi}^R(r)-{\bm k}^{-1}{\bm u}(r)\int\limits_R^r dr'\, {\bm \psi}^R(r'){\bf V}(r'){\bf \Psi}^R(r')
\nonumber \\
&-&{\bm k}^{-1}{\bm \psi}^R(r)\int\limits_r^\infty dr'\, {\bm u}(r'){\bf V}(r'){\bf \Psi}^R(r').
\label{LSEPsi^RgrR}
\end{eqnarray}
The asymptotic form of ${\bf \Psi}^R(r)$ as $r\to\infty$ can now easily be evaluated from the right hand side of this equation by
neglecting the last term,  since it goes to zero, and extending to infinity the upper limit of the integral in the second term.
{The result from this is given by}
\begin{equation}
{\bf \Psi}^R(r)\sim {\bm \psi}^R(r)-{\bm k}^{-1}{\bm u}(r)\int\limits_R^{\infty} dr'\, {\bm \psi}^R(r'){\bf V}(r'){\bf \Psi}^R(r').
\label{Psi^Ras}
\end{equation}
From this formula the final asymptotic form (\ref{MBC8+}) of ${\bf \Psi}^R(r)$ as $r\to\infty$ can be obtained by using (\ref{psigrR}).
{Therefore,} the scattering amplitude ${\bf {\cal A}}^R$ is given by
\begin{equation}
{\bf {\cal A}}^R={\bm A}^R-{\bm k}^{-1}\int\limits_R^\infty dr'\, \left[{\bf F}(r')+
{\bm u}(r'){\bm A}^R\right]{\bf V}(r'){\bf \Psi}^R(r').
\label{amplitudeA^R}
\end{equation}

The Green's function ${\bf G}^R(r,r')$ which satisfies the equation
\begin{eqnarray}
\left[{\bf H}_0 + {\bf L}(r) + {\bf C}^R(r)- {\bm k}^2\right]{\bf G}^R(r,r')\nonumber \\
+
{\bf V}^R(r){\bf G}^R(r,r')={\bf I}\delta(r-r')
\label{MSE^RG}
\end{eqnarray}
can also be defined with the help of Lippmann-Schwinger equation
\begin{equation}
{\bf G}^R(r,r')={\bm g}^R(r,r') - \int\limits_{R}^{\infty} dr'' \,{\bm g}^R(r,r''){\bf V}(r''){\bf G}^R(r'',r').
\label{lseG}
\end{equation}
This equation is also well defined due to (\ref{srpotV}) and has {a} unique solution. In the next subsection we use the
{asymptotics} of the Green's function ${\bf G}^R(r,r')$ in the special cases {where} $r'\le R$ and $r\gg R$. This
{asymptotics} can again be evaluated from the right hand side of the equation (\ref{lseG}) when $r>R$ and $r'\le R$
\begin{eqnarray}
{\bf G}^R&&(r,r')=
{\bm k}^{-1}{\bm u}(r){\bm \psi}^R(r')\nonumber \\
&&-
{\bm k}^{-1}{\bm u}(r)\int\limits_{R}^r dr''\, {\bm \psi}^R(r''){\bf V(r'')}{\bf G}^R(r'',r')
\nonumber \\
&&-{\bm k}^{-1}{\bm \psi}^R(r)\int\limits_{r}^\infty dr''\, {\bm u}(r''){\bf V(r'')}{\bf G}^R(r'',r')
\label{lseGG}
\end{eqnarray}
by neglecting the last term, since it goes to zero as $r\to\infty$,  and by extending the upper limit of integration to infinity
in the second term. {This gives the following expression}
\begin{eqnarray}
{\bf G}^R(r,r')\sim {\bm k}^{-1}{\bm u}(r) {\bf {\hat \Psi}}^R(r'),
\label{GPhi}
\end{eqnarray}
where
\begin{eqnarray}
{\bf {\hat \Psi}}^R(r')&=& {\bm \psi}^R(r')\nonumber \\
&-&\int\limits_{R}^\infty dr'' \, {\bm \psi}^R(r''){\bf V}(r''){\bf G}^R(r'',r').
\label{Phi}
\end{eqnarray}
{By direct calculations one can verify that the transposed matrix ${\bf {\hat \Psi}}^R(r)$ obeys the Lippmann-Schwinger equation
(\ref{LSE^R}) for ${\bf \Psi}^R(r)$ and, therefore, due to the uniqueness of the solution of this equation the following equality holds true}
\begin{equation}
{\bf {\hat \Psi}}^R(r)={{\bf \Psi}^R}^T(r),
\label{Phi=PsiT}
\end{equation}
where $T$ {is} the matrix transposition. The final form of the desired {asymptotics where} $r'\le R$ and $r\to\infty$ {are}
obtained by taking into account the representation (\ref{Psi^RleR}) and reads
\begin{eqnarray}
{\bf G}^R(r,r') \sim {\bm k}^{-1}{\bm u}(r) {{\bf a}^R}^T  {\bf {\hat j}}(r').
\label{GPhiPsi}
\end{eqnarray}

\subsection{The driven Schr\"{o}dinger equation and exterior complex scaling. The integral and the local representations
for the scattering amplitude.}\label{DSEECS}

{The solution ${\bf \Psi}^R$ just obtained can now be considered as the incoming wave. By its construction,} the action of
the operator ${\bf H}-{\bm k}^2$ on this incoming wave has the form
\begin{equation}
({\bf H}-{\bm k}^2){\bf \Psi}^R(r)=\left[{\bf C}_{R}(r)+{\bf V}_{R}(r)\right]{\bf \Psi}^{R}(r).
\label{H-k Psi^R}
\end{equation}
{If the "scattered" wave ${\bf \Psi}^{sc}$ is introduced by the expression}
\begin{equation}
{\bf \Psi}(r)= {\bf \Psi}^R(r)+{\bf \Psi}^{sc}(r),
\label{Psi S}
\end{equation}
then the Eq. (\ref{MSE}) for ${\bf \Psi}(r)$ transforms into the inhomogeneous equation for ${\bf \Psi}^{sc}(r)$
\begin{equation}
({\bf H}-{\bm k}^2){\bf \Psi}^{sc}(r)=-\left[{\bf C}_{R}(r)+{\bf V}_{R}(r)\right]{\bf \Psi}^R(r).
\label{predse}
\end{equation}
This equation {has two} key properties which are very important for application of the exterior complex scaling, i.e. the
inhomogeneous term vanishes outside of the radius $R$  and the solution  ${\bf \Psi}^{sc}$ has purely outgoing {asymptotics}. The final
form {for} the driven equation formulation can be obtained from (\ref{predse}) by using the following observation. The right
hand side term in (\ref{predse}) has the following explicit form
$$
\left[{\bf C}_{R}(r)+{\bf V}_{R}(r)\right]{\bf \Psi}^R(r)= \left[{\bf C}_{R}(r)+{\bf V}_{R}(r)\right]{\bf {\hat j}}(r){\bf a}^R.
$$
{By multiplying Eq. (\ref{predse}) by the inverse matrix $({{\bf a}^R})^{-1}$ from the right and then introducing the matrix}
\begin{equation}
{\bf \Phi}(r)={\bf \Psi}^{sc}(r)({{\bf a}^R})^{-1}
\end{equation}
the former equation transforms into
\begin{equation}
({\bf H}-{\bm k}^2){\bf \Phi}(r)=-\left[{\bf C}_{R}(r)+{\bf V}_{R}(r)\right]{\bf {\hat j}}(r).
\label{dse}
\end{equation}
The boundary conditions follow from (\ref{MBCzero}), (\ref{MBC8}) and (\ref{MBCzero+-}), (\ref{MBC8+}) 
\begin{eqnarray}
{\bf \Phi}(0)&=&0 \nonumber \\
{\bf \Phi}(r)&\sim& {\bm u}(r)({\bm A}^s-{\bf{\cal A}}^R)({{\bf a}^R})^{-1}, \ \ r\to \infty.
\label{Phibc}
\end{eqnarray}
Eqs. (\ref{dse},\ref{Phibc}) {provide} us with the final formulation {for} the driven Schr\"odinger equation.

The integral representation for the scattering amplitude which {results} from the driven Schr\"odinger equation formulation is the
{last feature discussed in this subsection.} In order to derive this representation  Eq. (\ref{dse}) should be recast into
\begin{eqnarray}
\left[{\bf H}_0+{\bf L}(r)+{\bf C}^R(r)+{\bf V}^R(r)-{\bm k}^2\right]{\bf \Phi}(r)= \nonumber
\\ -\left[{\bf C}_{R}(r)+{\bf V}_{R}(r)\right]\left[{\bf {\hat j}}(r)+{\bf \Phi}(r)\right].
\label{MSEphi}
\end{eqnarray}
Using the Green's function ${\bf G}^R(r,r')$ this equation can be rewritten in the integral form
\begin{eqnarray}
{\bf \Phi}(r) = - \int\limits_{0}^R dr'\,  {\bf G}^R(r,r')&&\left[{\bf C}_{R}(r')+{\bf V}_{R}(r')\right]
\nonumber \\
&& \times \left[{\bf {\hat j}}(r')+
{\bf \Phi}(r')\right].
\label{IntPhi}
\end{eqnarray}
The {asymptotics} of the solution ${\bf \Phi}(r)$ {follow} now from (\ref{IntPhi}) by taking into account  the {asymptotics}
(\ref{GPhiPsi}) of the Green's function  ${\bf G}^R(r,r')$. {This gives}
\begin{equation}
 {\bf \Phi}(r) \sim {\bm u}(r){{\bf a}^R}^T{\bm J}({\Phi}),
 \label{Phias}
 \end{equation}
\begin{equation}
{\bm J}({\Phi})= - {\bm k}^{-1}\int\limits_{0}^{R}dr'\, {\bf {\hat j}}(r')\left[{\bf C}(r')+{\bf V}(r')\right]
\left[{\bf {\hat j}}(r')+{\bf \Phi}(r')\right].
\label{APhi}
\end{equation}
By comparing (\ref{Phibc}) with (\ref{Phias}) and using the {definition given in} (\ref{totampl}) we obtain the final representation
for the total scattering amplitude $\bm A$
\begin{equation}
{\bm A}={\bm A}^C+{\bf {\cal A}}^R+ {{\bf a}^R}^T{\bm J}({\Phi}){\bf a}^R.
\label{totapmfin}
\end{equation}
Thus if the matrices ${\bf {\cal A}}^R$ and ${\bf a}^R$ have been calculated then the driven equation formulation provides {an}
alternative to the original formulation (\ref{MSE}), (\ref{MBCzero}) and (\ref{MBC8}). The important feature of this alternative formulation
is that in order to calculate the scattering amplitude ${\bm A}$ one needs to know ${\bf \Phi}(r)$ only in the finite interval $(0<r\le R)$.

The complex scaling application is based on the following arguments. The function ${\bf \Phi}(r)$ has purely outgoing wave {asymptotics}
at infinity. When the complex scaling transformation is applied to the boundary problem given by (\ref{dse}, \ref{Phibc}), the scaled boundary
problem will have the zero boundary conditions both at the origin and at infinity. The driving term in the driven Schr\"{o}dinger equations
(\ref{dse}) does not diverge at large distances under the complex scaling transformation of the coordinate {if, in the exterior
complex scaling,} a proper choice of the scaling point is made. {Therefore,} the necessary conditions for application of the exterior
complex scaling to the boundary problem given by (\ref{dse}, \ref{Phibc}) are fulfilled.
The formal scheme of this application is as follows. If we denote the complex scaled function as
$\tilde{{\bf \Phi}}(r)={\bf \Phi}(g_{\alpha,Q}(r))$ { then, for this function,} we obtain the boundary problem
\begin{eqnarray}
\label{SDSE}
\left\{
\begin{array}{l}
\left(\tilde{{\bf H}}-{\bm k}^2\right)\tilde{{\bf \Phi}}(r) = -\left[{\bf C}_{R}(r)+{\bf V}_{R}(r)\right]{\bf {\hat j}}(r)~,\\
\tilde{{\bf \Phi}}(0)  = 0~,\\
\tilde{{\bf \Phi}}(\infty) =0~,
\end{array}
\right.
\end{eqnarray}
where $\tilde{{\bf H}}$ represents the complex scaled Hamiltonian. The finite-range driving part remains unchanged after the complex scaling
if $R < Q$. {Furthermore}, if $R < Q$ then the complex scaling transformation does not change the value of the function  ${\bf \Phi}(r)$
in the region $r<R$  {and, as such,} ${\bm J}({\tilde {\Phi}})={\bm J}({{\Phi}})$. Thus, provided that we have solved the scaled problem
(\ref{SDSE}) for $\tilde{{\bf \Phi}}$, the scattering amplitude matrix ${\bm A}$ can be computed from the representation
\begin{eqnarray}
\label{intrepr1}
{\bm A} = {\bm A}^C + {\bf {\cal A}}^R + {{\bf a}^R}^T
{\bm J}({\tilde {\Phi}})
{\bf a}^R~.
\end{eqnarray}
The matrices ${\bf a}^R$ and ${\bf {\cal A}}^R$ defined by Eqs.~(\ref{aa^R}) and (\ref{amplitudeA^R}) have to be {determined} in order
to use (\ref{intrepr1}). According to their definitions, these matrices can be calculated if we know the function ${\bf \Psi}^R(r)$ in
the region $r>R$.  However, the numerical integration of the differential equation (\ref{MSE^R}) with an arbitrary potential ${\bf V}(r)$
and the point $R$ is a problem of {similar} complexity {to} the initial scattering problem (\ref{MSE}-\ref{MBC8}).
{Therefore, in practical calculations,} we choose the point $R$ to be large enough in order to assume  that ${\bf V}^R(r) = 0$.
The validity of this assumption should be checked for each potential under investigation. The truncation of potentials decreasing faster
than $r^{-2}$ at infinity does not lead to principal errors contrary to the truncation of the Coulomb potential. In section \ref{HN}
we will analyze how the truncation of the potential ${\bf V}$ affects the total cross section.

If we {set} ${\bf V}^R=0$, then Eqs.~(\ref{aa^R}) and (\ref{amplitudeA^R}) yield
\begin{eqnarray}
\label{a12+a120} {\bf a}^{R} &=& {\bm a}^R~,\\
\label{A12+A120}   {\bf {\cal A}}^R &=& {\bm A}^R~,
\end{eqnarray}
where the matrices ${\bm a}^R$ and ${\bm A}^R$ are diagonal. The integral representation for the scattering amplitude (\ref{intrepr1})
{then} transforms into
\begin{eqnarray}\label{intrepr2}
{\bm A} = {\bm A}^C +{\bm A}^R + {\bm a}^R
{\bm J}(\tilde{{\bf \Phi}})
{\bm a}^R .
\end{eqnarray}
Furthermore, the asymptotic relation~(\ref{Phias}) becomes exact for $r\ge R$. Taking into {account the fact that} the function
at the point $r=R$ is not complex scaled, $\tilde{{\bf \Phi}}(R) = {\bf \Phi}(R)$, we conclude that
\begin{eqnarray}\label{local}
{\bm A} = {\bm A}^C+{\bm A}^R + {\bm u}^{-1}(R)\tilde{{\bf \Phi}}(R)\,{\bm a}^R ~.
\end{eqnarray}
The last expression provides us with the local representation for the scattering amplitude {which is an} alternative to the integral
representation~(\ref{intrepr2}).

\section{Numerical approach, results and discussions}\label{Num}

The equation with zero boundary conditions (\ref{SDSE}) {together with the} two alternative representations for the scattering
amplitude (\ref{intrepr2}) and (\ref{local}) can be directly implemented numerically. However, the derived equations can be slightly
modified in order to {obtain a more numerically stable} implementation for large orbital momentum.  Calculations with such momenta
are necessary to {achieve} converged results for the total cross sections, see section~\ref{HN}.
In the section \ref{ansolv} {we described a method for} constructing the solution to the diagonal part of the equation (\ref{MSE})
which incorporates the long range Coulomb interaction for $r>R$.
Another approach is obtained through splitting the centrifugal term ${\bf L}(r)$ for each partial wave,
\begin{equation}
{\bf L}(r)= {\bf L}_{R}(r)+{\bf L}^R(r)
\label{L-split}
\end{equation}
in the same way as is done for the Coulomb interaction.
Eq. (\ref{MSElong}) in this case transforms into
\begin{equation}
\left[{\bf H}_0 + {\bf L}^R(r) +{\bf C}^R(r) - {\bm k}^2\right]{\bm \psi}^{R}(r)=0.
\label{MSElong+L}
\end{equation}
functions ${\hat h}^{\pm}_{\ell}(k_n r)$ should be replaced by {the} trigonometric functions $\sin(k_n r)$ and $e^{\pm i k_n r}$,
respectively, {in all of the formulas given in} the section \ref{ansolv}.
Then the final equation (\ref{SDSE}) transforms into
\begin{eqnarray}
\label{SDSE1}
\left\{
\begin{array}{l}
\left(\tilde{{\bf H}}-{\bm k}^2\right)\tilde{{\bf \Phi}}(r) = -\left[{{\bf C}}_R(r)+{{\bf L}}_R(r)+{{\bf V}}_R(r)\right]{\bf sin}(r)~,\\
\tilde{{\bf \Phi}}(0)  = 0~,\\
\tilde{{\bf \Phi}}(\infty) =0~,
\end{array}
\right.
\end{eqnarray}
where $[{\bf sin}(r)]_{nm} = \delta_{nm}\sin(k_nr) $.
The scattering amplitude matrix {can then be} calculated with the same local
(\ref{local}) and integral (\ref{intrepr2}) representations.
However, the matrices ${\bf a}^R$ and ${\bf {\cal A}}^R$  must be calculated using the expressions (\ref{a012}) and (\ref{A012})
{with the above stated modifications}.

The boundary problem (\ref{SDSE1}) rather than (\ref{SDSE}) is used in {all of the} calculations reported in this section.
For the sake of completeness, it should be noted that there exists {a} third possibility, where the unperturbed Hamiltonian also
includes the Coulomb potential.
Then the solution of the scattering problem with the long range diagonal part of the potential is explicitly
given in terms of the Coulomb functions.
This version has been explored e.g. in papers~\cite{McCurdy2004,Kruppa}.
However, it is not clear how this technique can be extended to systems consisting of three or more particles.
Therefore, we do not focus on this option {in this paper}.

We use the derived equations in order to study two simple models.
{We consider first} the short-range one-channel Noro-Taylor
potential~\cite{NoroTaylor80} supplied with the Coulomb interaction.
The second example is the more realistic two-channel model for the $N^{3+} + H \to N^{2+} + H^{+}$ reaction~\cite{REBB84,ksh_nh3}.
This latter model is composed of the Coulomb interaction and molecular inverse power potentials.

As the numerical method for the solution of Eq.~(\ref{SDSE1}), we have chosen the FEM-DVR approach described in~\cite{RC00}.
This approach can be considered as a finite element method (FEM) with a special choice of the basis functions on each element,
namely the polynomial Lobatto shape functions~\cite{RC00}. With this choice, the matrix elements of local operators (i.e. potentials)
are approximately diagonal with respect to the basis functions. The error introduced by such approximation does not influence the convergence
rate of the FEM. The parameters of the numerical applications were chosen such that the numerical inaccuracies were {negligible.}

{Although} the radius $R$ and the exterior complex rotation radius $Q$ are allowed to be different in the scheme described above, we have
not found any advantages {to} keeping them distinct.
Hence, in our calculations we choose $Q=R$.
The preliminary calculations have also confirmed that the specific choice of the exterior complex
scaling defined by Eqs.~(\ref{CS}) and (\ref{ECS}) does not affect the results.
Therefore, we have used the sharp exterior complex scaling~\cite{S79} {in the calculations}.
\begin{equation}
\label{ECS_sharp}
g_{\alpha,R}(r) = \left\{
\begin{array}{lll}
r                       & \mbox{for }& r \le R \\
R+(r-R) e^{\imath \alpha} & \mbox{for }& r > R
\end{array}
\right. .
\end{equation}
In contrast {to applications of the complex scaling method to computing resonances}, the choice of the scaling angle $\alpha$ is
not limited here by any additional restrictions.
Therefore, the angle has been chosen to be close to $90$ degrees to enable the fastest decay of the wave function at infinity.

In order to employ the boundary conditions in Eq. (\ref{SDSE1}) into the numerical scheme, we introduce the maximal radius $R_{max} > R$,
where the second boundary condition of (\ref{SDSE1}) is implemented. The radius $R_{max}$ should be considerably larger than $R$, such that
the wave function decays on the interval $[R,R_{max}]$. As soon as $R_{max}$ {is sufficiently large}, no noticeably errors in
the results {are observed}.

\subsection{Specific aspects   
of the calculations with the Coulomb potential}\label{Noro}

{Consider the scattering problem} on the one-channel Noro-Taylor potential~\cite{NoroTaylor80} in the presence of the repulsive
Coulomb interaction
\begin{equation}
\label{pot-example}
 V(r) = 15 r^2 e^{-r}, \qquad C(r) = 2/r.
\end{equation}
{Channel indices are not used since the model contains only one channel}. The reduced mass was chosen to be $\mu=1$. The total
cross section for this potential with the Coulomb tail is infinite but one can analyze the partial wave cross sections.
In the computations, we have used 1000 finite elements with Lobatto polynomials of sixth degree.

\begin{figure}[t]
\centering
\includegraphics[scale=0.8]{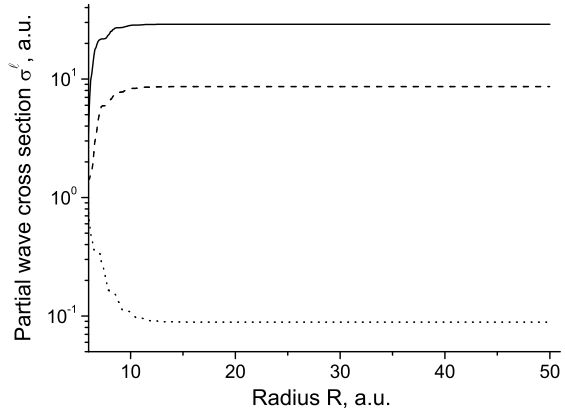}
\caption{The partial wave cross sections $\sigma^0$, $\sigma^5$, and $\sigma^{10}$ (the dotted, dashed and solid lines, respectively),
are plotted {as a function} of the radius $R$ for the energy $E=3$ a.u.}
\label{F:NT-R}
\end{figure}
Let us first discuss the influence of the radius $R$ on the partial wave cross sections.  In Fig.~\ref{F:NT-R} we plot the partial
wave cross sections {as a function of} the radius $R$.
As the short-range part of {the} potential~(\ref{pot-example}) decreases very fast, an accurate value for the cross section
can be obtained already for rather small radii $R$ starting from $R=20$ a.u. These values depend on the energy $E$ and on the short
range part of the potential such that $V^s(R) \ll E$. The accuracy is not influenced by the value of the angular momentum $\ell$,
at least {for moderate} values of $\ell$.

The matrices ${\bm a}^R$ and ${\bf A}^R$ are defined through the explicit expressions (\ref{a012}) and (\ref{A012}).
The Coulomb wave functions are used in Eq.~(\ref{a012}) and in the local representation (\ref{local}). Keeping in mind the three-body
generalization of our approach~\cite{EVLSMYY09}, {we can check here whether} we can completely avoid using the Coulomb functions
in our calculations.

The standard approach for {this check} is {to use} the asymptotic expansion of the Coulomb functions for large $R$. One can
then show (see for example~\cite{AbrSt} for the one-channel case) that two first terms in the $1/R$, $R\to \infty$, expansion for
${\bm u}(R)$ and ${\bm a}^R$ are given by
\begin{eqnarray}
\label{firstorder}
[{\bm u}(R)]_{mn} &\sim& \delta_{mn}(1+u_n)\exp[ i\theta_n]~,\nonumber\\
\left[{{\bm a}^R}\right]_{mn} &\sim&\delta_{mn}{\left(1 - \omega_n\right)}
  \exp\left[i \eta_n\log(2k_nR)\right]~.
\end{eqnarray}
Here $\theta_n = k_nR -\pi\ell/2 - \eta_n\log(2k_nR)$, and
\begin{eqnarray}
\label{parameter}
u_n &=& \frac{\eta_n+i\left(\ell(\ell+1)+\eta_n^2\right)}{2k_nR}~,
\nonumber\\
\omega_n &=& \frac{i\eta_n^2 +\eta_n\exp\left[2i k_nR - i\pi\ell \right]}{2k_nR}~.
\end{eqnarray}
In the numerical scheme (\ref{SDSE1}), the {similar asymptotics are} modified to be equal to
\begin{eqnarray}
\label{zeroorder}
[{\bm u}(R)]_{mn} &\sim& \delta_{mn} \exp\left[i\theta_n\right]~,
\nonumber\\
\left[{\bm a}^{R}\right]_{mn}  &\sim& \delta_{mn}
\exp\left[i \left( \eta_n\log(2k_nR) +\pi\ell/2 \right)\right]~,
\end{eqnarray}
for the main terms of {the asymptotics}. The next order terms are given by
\begin{eqnarray}
\label{firstorder_mod}
\left[{{\bm a}^R}\right]_{mn} &\sim&\delta_{mn}{\left(1 - \omega_n\right)}
  \exp\left[i \left( \eta_n\log(2k_nR) +\pi\ell/2 \right)\right]~, \nonumber\\
\omega_n &=&\frac{i\eta_n^2 + \eta_n\exp\left[2i k_nR\right]+i\ell(\ell+1)}{2k_nR}~,
\end{eqnarray}
while ${\bm u}(R)$ coincides with that in Eq.~(\ref{parameter}). {It is noted} that the expressions (\ref{zeroorder}),
(\ref{firstorder}), and  (\ref{firstorder_mod}) are only valid when $|u_n|\ll 1$.
\begin{figure}[t]
\centering
\includegraphics[scale=0.8]{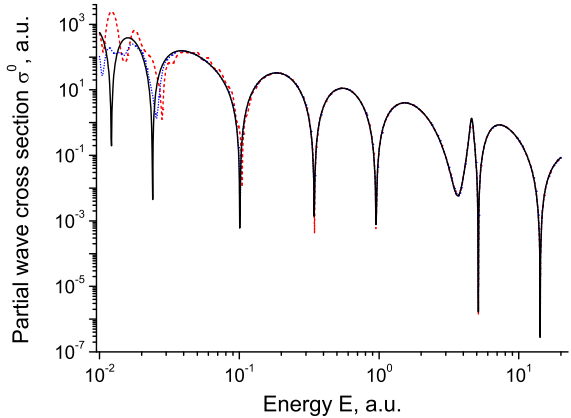}
\caption{(Color online) The partial-wave cross section $\sigma^0$ {as a} function of the scattering energy $E$. The results
for the exact boundary condition~(\ref{a012}) (the solid line), asymptotic boundary condition~(\ref{zeroorder}) (the dashed line),
and {the} asymptotic boundary condition with the correction term~(\ref{firstorder_mod}) (the dotted line) are shown.
The radius $R=100$ a.u.}
\label{F:NT-partial0}
\end{figure}
\begin{figure}[t]
\centering
\includegraphics[scale=0.8]{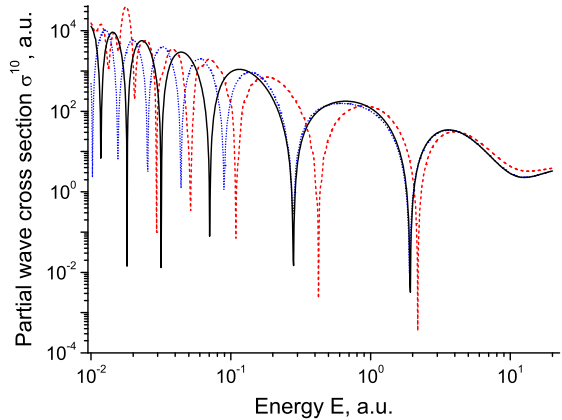}
\caption{(Color online) {As in Fig.~\ref{F:NT-partial0} but for the partial-wave cross section $\sigma^{10}$.}}
\label{F:NT-partial10}
\end{figure}
Let us check how the derived {asymptotic representations} influence the cross section calculations.
In Figs.~\ref{F:NT-partial0} and~\ref{F:NT-partial10}
we plot the partial wave cross sections as {a function} of the scattering energy $E$. We compare the cross sections obtained with three
different boundary conditions at the point $r=R$, namely the exact boundary condition~(\ref{a012}), the asymptotic boundary
condition~(\ref{zeroorder}), and the asymptotic boundary condition with the correction term~(\ref{firstorder_mod}).  For the chosen parameters
and energy regions, the results for the integral and local amplitude representations are indistinguishable {and, as such, we only plot
the results for the integral representation}. For zero angular momentum, Fig.~\ref{F:NT-partial0}, all curves practically coincide.
Some differences appear {only} for small energies, where the value for the asymptotic parameter $u$ given by Eq.~(\ref{parameter})
approaches $|u| \approx 0.25$ from below for $E=0.03$ a.u. and gets even bigger for smaller energies. For the momentum $\ell=10$,
Fig.~\ref{F:NT-partial10}, the difference between the boundary condition~(\ref{a012}) and the two other boundary conditions
is {more pronounced and clearly increases with decreasing energy.} {Conversely}, the results for the exact boundary
conditions~(\ref{a012}) and the asymptotic boundary condition with the correction term (\ref{firstorder_mod}) agree quite well,
{even starting from relatively small energies} $E \geq 0.2$ a.u. . On the contrary, the results for the simple
{asymptotics}~(\ref{zeroorder}) disagree with the correct results {over} the entire energy region shown.
Thus, {we conclude} that the correction term introduced in (\ref{firstorder_mod}) improves the cross section essentially
when compared to the simple asymptotics~(\ref{zeroorder}).
This also means that with the correction term (\ref{firstorder_mod})
{we can, for a given energy,} use a smaller value of $R$ in order to reach the same accuracy.

The accuracy of the correction depends on the parameters $u_n$ defined in Eq.~(\ref{parameter}).
If the angular momentum is fixed, the accuracy is improved when $kR$ increases.
{Conversely}, for chosen scattering energy and radius $R$, the accuracy gets
worse when the angular momentum increases.

{With respect to the differences between} the integral~(\ref{intrepr2}) and local~(\ref{local}) representations of the scattering
amplitudes, they seem to depend on the chosen boundary conditions.
For the exact boundary condition~(\ref{a012}) both these representations give identical values, while they result in different values for
the asymptotic boundary conditions~(\ref{zeroorder}, \ref{firstorder_mod}). This difference grows when $kR$ gets smaller, as discussed
in Ref.~\cite{VEYY09}.
%


Our approach is rigorous in treating the Coulomb interaction both in {the} one channel case \cite{VEYY09} and in the multi
channel {case described in} the present paper. {The one} channel scattering problem with {a} potential that
decreases at large separations faster than $r^{-2}$ was treated in the paper \cite{RBBC97}. The driven equation of the form (\ref{SDSE})
was used and then solved numerically. In the notations of this paper, the scattering amplitude of Ref.~\cite{RBBC97} is expressed as
\begin{equation} \label{intreprold}
{\bm A} =  {\bm J}(\tilde{{\bf \Phi}})~.
\end{equation}

Let us compare this representation with the exact one~(\ref{intrepr2}).
First of all, if the Coulomb interaction is not present in the system, then ${\bf A}^C=0$ and ${\bm a}^R={\bf I}$, ${\bm A}^R =0$.
Hence, the representations (\ref{intrepr2}) and (\ref{intreprold}) are identical.
If the Coulomb interaction is present, we can use the fact that the matrices ${\bm a}^R$ and ${\bm k}$ are diagonal and {so} rewrite
the exact matrix representation (\ref{intrepr2}) in terms of the matrix elements as
\begin{equation}
\label{intrepr3}
A_{fi} = A^C_f \delta_{fi}+ A^R_{f}\delta_{fi}
+  a^R_{f} {J(\Phi)}_{fi}\,a^R_{i} ~.
\end{equation}
%
%
The diagonal elements $A^C_{f}$ and $A^R_{f}$ are equal to zero for the channels without the Coulomb interaction.
{Therefore, for all inelastic transitions and for elastic transitions within the non-Coulomb channels, we obtain}
\begin{equation}
\label{intrepr4}
A_{fi} = a^R_{f}{J(\Phi)}_{fi}\,a^R_{i} ~.
\end{equation}
If the Coulomb potential is absent in both {the} $f$ and $i$ channels, then $a^R_f = a^R_i =1$, {and} there is no difference
between {the} two representations. If it is present in one or both {of the} inelastic channels {then, according
to the asymptotic representation} (\ref{firstorder}), $|a^R_{k}| \to 1$ when $R\to \infty$.
This means that the representation {described in} (\ref{intreprold}) gives the same cross section as the exact representation
(\ref{intrepr2}) provided that $R$ is {sufficiently large}. {However, it should be noted that, as opposed to the cross section,}
the scattering amplitude is not calculated correctly even in the $R\to \infty$ limit.

\begin{figure}[t]
\centering
\includegraphics[scale=0.8]{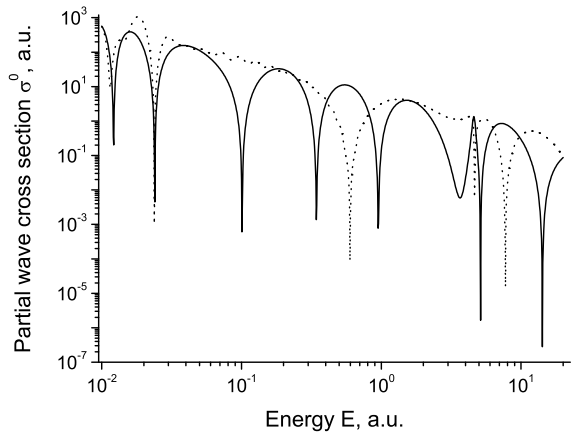}
\caption{The $s$-wave cross section $\sigma^0$ {as a function} of the scattering energy $E$. The cross-section {is} calculated
using the exact integral representation (\ref{intrepr2}) (the solid line) and the representation {given in} (\ref{intreprold})
(the dotted line).}
\label{F:NT-rescigno}
\end{figure}

{In summary of this comparison}, we can distinguish three different combinations of channels $f$ and $i$ for the Coulomb multi
channel scattering problem. {If the Coulomb interaction is absent in both channels }(non-Coulomb elastic and inelastic channels),
the results given by the {representation shown in} (\ref{intreprold}) are identical with the exact values defined with Eq.~(\ref{intrepr2}).
If the Coulomb interaction is present in an inelastic channel, the partial cross sections calculated with Eq.~(\ref{intreprold}) approach the
correct values as $R\to \infty$.
For the Coulomb elastic channels, the representation~(\ref{intreprold}) fails to give the correct answer. In order to illustrate the latter
statement, we compare in Fig.~\ref{F:NT-rescigno} the $s$-wave cross section calculated with both the exact integral representation~(\ref{intrepr2})
and the scattering amplitude~(\ref{intreprold}) for the potential~(\ref{pot-example}).
{These cross sections completely disagree.}

\subsection{The $N^{3+} + H \to N^{2+} + H^{+}$ reaction}\label{HN}

{Here we study the two channel charge transfer $ N^{3+}(1s^22s^2) + H(1s) \to NH^{3+} \to N^{2+}(1s^22s^23s) + H^+ $.}
The matrix potential describing this reaction is parameterized as
\begin{eqnarray}
\label{HN3_pot}
{\bf V}(r) &=& 2\mu \left(
  \begin{array}{cc}
    4000 r^{-8} - 20.25 r^{-4} & 0.5 r^2 e^{-r} \\
    0.5 r^2 e^{-r} & -0.235\\
  \end{array}
\right)~,
\nonumber \\
{\bf C}(r) &=& 2\mu \left(
  \begin{array}{cc}
    0 & 0 \\
    0 & 2 r^{-1}\\
  \end{array}
\right)~.
\end{eqnarray}
The reduced mass was taken to be $\mu = 1713.5$~a.u.
These parameters are taken from~\cite{REBB84} while motivation for the choice
of the model and its parameters are found in~\cite{BBER83}.
In the numerical study of this system we have also used the FEM-DVR with the sixth degree Lobatto polynomials.
The number of equidistant finite elements depends on the radius $R$.
Their density has been chosen to be 12 elements per atomic unit.

\begin{figure}[t]
\centering
\includegraphics[scale=0.8]{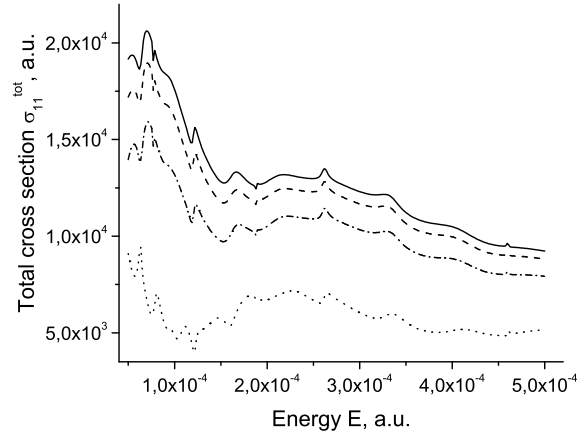}
\caption{The total cross section $\sigma^{tot}_{11}$ as {a} function of the energy $E$.
The values for the radii $R=100$~a.u. (the solid line), $R = 70$~a.u. (the dashed line), $R = 50$~a.u.
(the dash-dotted line) and $R=30$~a.u. (the dotted line) are plotted.}
\label{F:HN3-total}
\end{figure}

In this {discussion} we will mainly concentrate on the {development} of the method {to determine an} appropriate
choice of the only parameter {in} our approach, namely the cut-off radius $R$. In order to show the importance of {this}
choice, we plot in Fig.~\ref{F:HN3-total} the total elastic cross section, $\sigma^{tot}_{11}$ (\ref{TCS}), for different values of the
radius $R$. {Analysis of these results show that although the form of the cross-sections are similar, calculations with relatively
small values of $R$ essentially underestimate the value of the cross section compared to those calculated with the largest value of $R$.}
Such a situation can easily result in {large absolute errors in the computation of the the total cross section}.
The total rearrangement cross section, $\sigma^{tot}_{12}$, is plotted {in} Fig.~\ref{F:HN3-inter-total}. In this channel, the total
cross section converges already at the relatively small value of the cut off radius, $R=50$ a.u. Hence we expect that the $R$ value which
guarantees the convergence depends on the behavior of the channel potentials at large distances.

\begin{figure}[t]
\centering
\includegraphics[scale=0.8]{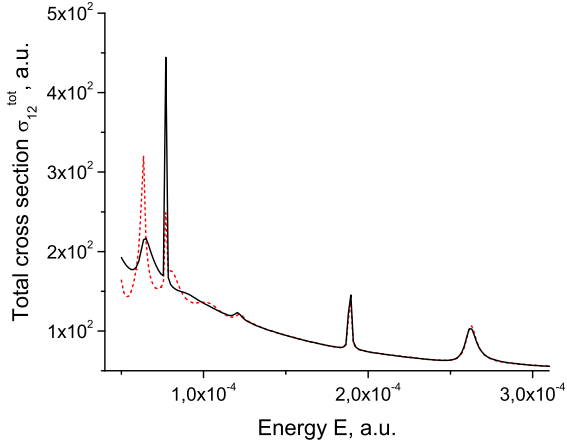}
\caption{(Color online)The total cross section $\sigma^{tot}_{12}$
as the function of the energy $E$. The values for the radii $R=50$~a.u. (the solid line) and $R=30$~a.u. (the dashed line) are plotted.}
\label{F:HN3-inter-total}
\end{figure}

Let us therefore estimate the error which is introduced in the total cross section (\ref{TCS}) due to {the choice of} potential cut-off.
First of all, the cut-off results in {a change} of each partial cross section.
In order to estimate this change, we should refer
to the representation of the scattering amplitude (\ref{totapmfin}) for the uncut potential.
Assuming that the change of amplitude due to
cutting the potential is much smaller than the amplitude itself, we can see that the main change comes from the last term in Eq.~(\ref{totapmfin}).
{Therefore, for large $R$,} we have
\begin{equation}  \label{ampl_perturb}
\delta {\bm A} \approx - {{\bf W}^R}^T {\bf a}^R {\bm J}({\Phi}) {\bf a}^R
- {\bf a}^R {\bm J}({\Phi}) {\bf a}^R {\bf W}^R,
\end{equation}
where
\begin{equation}  \label{Qint}
{\bf W}^R =
 {\bm k}^{-1}\int\limits_R^\infty  d r'\, {\bm u}(r'){\bf V}(r'){\bf \Psi}^R(r') .
\end{equation}
Using the asymptotic expansions for the functions ${\bm u}(r)$ and ${\bf \Psi}^R(r)$ for large $r\ge R$, {for the matrix elements
of the integral we obtain}
\begin{equation}  \label{Qint_as}
{W}^R_{mn} \approx
\int\limits_R^\infty  d r'\,
{k}^{-1}_{m} e^{i (k_m r' -\pi\ell/2)} V_{mn}(r') \sin{(k_n r' -\pi\ell/2)} .
\end{equation}
Let us assume that $dV_{mn}(r)/dr \ll V_{mn}(r)$ for large $r$.  {For example, such an assumption is valid for the important type
of potentials: inverse power potentials.}
Integrating (\ref{Qint_as}) by parts, we find for the main term of the ${W}^R_{mn}$ {asymptotics} at large $R$
\begin{equation}  \label{Qint_calc1}
{W}^R_{mn} \approx \frac{1}{2} {k}^{-1}_{m} V_{mn}(R)
\left[ \frac{e^{i(k_m-k_n)R}}{k_n-k_m} + \frac{e^{i(k_m+k_n)R -i\pi\ell}}{k_n+k_m}
\right]
\end{equation}
for $k_n \neq k_m$, and
\begin{equation}  \label{Qint_calc2}
{W}^R_{mn} \approx \frac{1}{2} {k}^{-1}_{m} V_{mn}(R)
\left[ \frac{i R}{\beta - 1} + \frac{e^{2ik_m R -i\pi\ell}}{2 k_m}
\right]
\end{equation}
for $k_n = k_m$, and {where} the potential $V_{mn}(r)$ {decreases} as $\sim C/r^\beta$ at infinity. For the sake of simplicity,
we omit here the indexes $mn$ for the parameter $\beta$.
These expressions result in the components of the amplitude change $\delta {\bm A}$
\begin{equation}  \label{ampl_perturb_comp}
\left[\delta {\bm A}\right]_{fi} \approx
- \sum_n \left( W^R_{nf} J_{ni}({\Phi}) a_n^R a_i^R + J_{fn}({\Phi}) a_f^R a_n^R W^R_{ni} \right).
\end{equation}
The corresponding change in the partial cross section (\ref{PCS}) {is given by}
\begin{equation} \label{crossec_perturb}
\delta {\sigma}^{\ell}_{fi}  \leq \sqrt{{\sigma}^{\ell}_{fi}} k_i
\sqrt{\frac{k_i}{k_f} \frac{1}{\pi(2\ell+1)}} \ |\left[\delta {\bm A}\right]_{fi}|~.
\end{equation}

The representation (\ref{ampl_perturb_comp}) shows that, in general, the amplitude error due to the potential cut-off depends on the amplitude
in {the} various channels. For a specific system, however, some channel potentials can decrease much faster than others, resulting in a
simpler description. For example, in the system~(\ref{HN3_pot}) {the off-diagonal potentials decrease exponentially and this also means
an exponential decrease in $W^R_{12}$ and $W^R_{21}$.} {Thus, for the elastic non-Coulomb channel $1 \to 1$,} we find that $a_1^R=1$ and
\begin{equation}  \label{ampl_11}
\left[\delta {\bm A}\right]_{11} \approx - 2 J_{11}({\Phi}) W^R_{11}.
\end{equation}
Taking into account the {asymptotics} (\ref{Qint_calc2}), {we obtain} for the relative error of the partial cross section
\begin{equation} \label{crossec_11}
\frac{\delta {\sigma}^{\ell}_{11}} {{\sigma}^{\ell}_{11}}  \leq
\frac{1}{3\sqrt{\pi(2\ell+1)}} \ R |V_{11}(R)|
\end{equation}
for large $R$. {In this case, the relative error decreases as} $\sim R^{-3}$. The absolute error of the total cross section is estimated as
\begin{eqnarray} \label{crossec_11tot}
\delta {\sigma}^{tot}_{11} = \sum_{\ell=0}^\infty \delta {\sigma}^{\ell}_{11}
&\leq&
R |V_{11}(R)| \sum_{\ell=0}^\infty \frac{1}{3\sqrt{\pi(2\ell+1)}} \ {\sigma}^{\ell}_{11}
\nonumber \\
&\leq&  \frac{R |V_{11}(R)|} {3\sqrt{\pi}} \ {\sigma}^{tot}_{11} .
\end{eqnarray}
{The absolute error also decreases with the same rate $\sim R^{-3}$ for large $R$ and its value can be easily controlled in the calculations}.

Comparing the results {plotted in} Fig.~\ref{F:HN3-total} and given by the estimation (\ref{crossec_11tot}), one can see that
Eq.~(\ref{crossec_11tot}) essentially underestimates the error.  In order to clarify this difference, let us discuss the dependence
of the partial-wave cross section $\sigma_{11}^\ell$ on the orbital momentum $l$. The corresponding values are plotted in Fig.~\ref{F:HN3-l}
for {a few values of the} cut-off radii. The calculation for $R=500$~a.u. can be considered, in the range of the figure,
{as a reflection of} the exact results. {Analysis of the results shows that there are three distinct regions of $\ell$-dependence}.
For {small values of} $\ell$, this dependence is not regular. For medium values of $\ell$ one can see a region of linear dependence that
results in the inverse power behavior of the cross section {as a function $\ell$}. Finally, for large values of $\ell$, the partial wave
cross sections decrease very fast.
So while the results for the values of $\ell$ in the medium region can be considered as good approximations to the correct cross sections, the
calculation for the large $\ell$ {fails completely}.

\begin{figure}[t]
\centering
\includegraphics[scale=0.8]{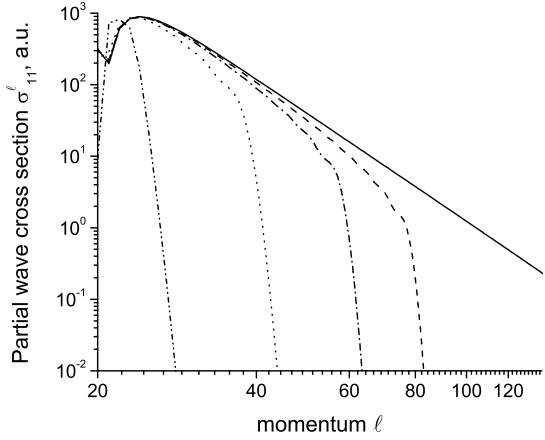}
\caption{The partial wave cross section $\sigma_{11}^\ell$ as the function of the momentum $\ell$ for the energy $E=2\cdot 10^{-5}$~a.u.
is plotted for different values of $R$: $R=500$~a.u. (the solid line), $R=100$~a.u. (the dashed line), $R=75$~a.u. (the dash-dotted line),
$R=50$~a.u. (the dotted line) and $R=30$~a.u. (the dashed line).}
\label{F:HN3-l}
\end{figure}

In order to analyze Fig.~\ref{F:HN3-l}, we should recall the amplitude definition (\ref{APhi}).
For a {given} cut-off radius $R$ and {large $\ell$,} such that $k_n R \leq \ell$, one can use the {asymptotics} of the Riccati-Bessel function with respect to
its index~\cite{AbrSt}:
\begin{equation} \label{riccati-asymp}
{\hat j}_\ell(r) \sim \frac{1}{\sqrt{2e}} \left( \frac{e k_n r}{2\ell} \right)^{\ell+1},
\quad r \le R .
\end{equation}
This function decays extremely fast, so does the integral ${\bm J}({\Phi})$. {However}, the exact scattering amplitude for the uncut
potential {has different behavior}. In order to find its {asymptotics}, {we notice} that the expression (\ref{ampl_perturb})
does not describe the leading asymptotic term as the condition $\delta {\bm A} \ll {\bm A}$
{is no longer satisfied because of the fast decrease}
of ${\bm J}({\Phi})$. {Conversely,} the leading asymptotic term is now the integral term in the representation (\ref{amplitudeA^R})
for ${\bf {\cal A}}^R$:
\begin{equation} \label{ampl-semiclas}
{\cal A}^R_{mn} \approx
- \int\limits_R^\infty  d r'\, {k}^{-1}_{m} {\hat j}_\ell{(k_m r')} V_{mn}(r') {\hat j}_\ell{(k_n r')}.
\end{equation}
For the $1 \to 1$ channel, the potential $V_{11}$ consists of a superposition of the inverse power potentials $1/r^\beta$~(\ref{HN3_pot}). The integral with
$1/r^\beta$ standing instead of $V_{11}$ can  be evaluated  explicitly~\cite{Gradshtein}.  This leads to the following asymptotics
of the partial wave cross section ~(\ref{PCS}) at the large values of the momentum $\ell$  in the case of the inverse power potential
%
\begin{equation} \label{sigmal_invp}
{\sigma}^\ell_{11} \approx
C_\beta {k_1^{2\beta-6}} \ell^{3-2\beta}.
\end{equation}
The constant $C_\beta$ is known explicitly and depends {only on the potential}.
As the $V_{11}$ component of the potential~(\ref{HN3_pot}) decreases as $\sim 1/r^4$, this results in the $\sigma^\ell_{11} \sim 1/{\ell^5}$
asymptotic behavior of the partial wave cross section.
{This behavior} explains the linear intermediate region in the partial wave cross sections in Fig.~\ref{F:HN3-l}, {especially so}
for the semi-exact results with $R=500$~a.u.

The numerical results, however, are computed with the cut-off potential so the amplitude (\ref{ampl-semiclas}) is not taken into account.
{So when $\ell$ becomes sufficiently large that the behavior of the function ${\hat {j_\ell}}(r)$ approaches
the asymptotics~(\ref{riccati-asymp}), the behavior of the calculated partial cross section changes}. {They vanish very fast
when compared} to the correct values, {and this} can be seen in Fig.~\ref{F:HN3-l}. When summing up the partial wave cross sections
{to obtain the total cross sections}, this might {give rise to the} erroneous impression that the total cross section
is already converged. This implies that such incorrect behavior can influence the accuracy of {the calculated total cross section}.
Furthermore, we cannot improve the accuracy by increasing the number of partial wave cross sections taken into account {since}
the higher partial-wave cross sections are not correctly calculated.

Let us estimate the sum of discarded partial wave cross sections for the inverse power potential $1/r^{\beta}$.
We denote by $\ell_R \sim e k_1 R/ 2$ the value of $\ell$ when the
{asymptotics}~(\ref{riccati-asymp}) should be taken into account.
The correction to the total cross section $\Delta {\sigma}^{tot}_{11}$ can then be estimated by using (\ref{sigmal_invp}) as
\begin{equation} \label{sigmaerr1}
\Delta {\sigma}^{tot}_{11} = \sum_{\ell=\ell_R}^\infty \sigma^\ell_{11}
\approx C_\beta {k_1^{2\beta-6}} \sum_{\ell=\ell_R}^\infty \ell^{3-2\beta}
\approx \sigma^{\ell_R}_{11} \frac{\ell_R}{2(\beta-2)} .
\end{equation}
For a given inaccuracy in the total cross section $\Delta {\sigma}^{tot}_{11}$, the latter estimation can be considered as the equation
for the minimal value of $R$ which guarantees the {requested} accuracy.
According to the representation~(\ref{totapmfin}), the amplitude inaccuracies (\ref{ampl_perturb_comp}) and (\ref{ampl-semiclas}) are to
be summed in order to {obtain} the total estimation. This implies that the double sum of partial wave cross sections (\ref{crossec_11})
and (\ref{sigmal_invp}) gives the estimate for the inaccuracy in the cross section.

We would like to stress that {the} estimates~(\ref{crossec_11tot}, \ref{sigmaerr1}) are specific not only to 
our approach. Any numerical
method which cuts the potential in {one way or another}, experiences these errors. This implies that estimates similar to
Eqs.~(\ref{crossec_11tot}, \ref{sigmaerr1}) should be adopted in order to guarantee the accuracy of the numerical calculations.
Finally, we have compared our two-body partial wave and total cross section results with {those} described in ref.~\cite{ksh_nh3}
using a {log-derivative} method.  We find an agreement within the accuracy discussed. We also find that the needed computational
resources are comparable.

\section{Summary}\label{sum}

Inspired by the work of Nuttall and Cohen~\cite{NutCoh} and Rescigno {\it et al}~\cite{RBBC97} we have developed 
a theory which enables the calculation of two-body multi channel charged particle scattering.
{This theory is formulated with the aim of being able to
generalize from two-body to three-body charged particle multi channel quantum scattering, as briefly outlined in~\cite{EVLSMYY09}}.

The entire potential is split into a core and a tail potential. The scattering problem {in the first step} is solved for the diagonal,
analytically solvable part of the tail potential.
The solutions of this problem are then used to construct the corresponding Green's function.
{These are then used to derive the Lippmann-Schwinger equation, the solution of which is the wave function corresponding to the entire tail potential, including the off-diagonal elements}.
This {resulting} wave function is then used as an incident wave when formulating a driven Schr\"{o}dinger equation, which
has the desired scattering function as a solution. Using exterior complex scaling, this scattering wave function can be obtained from a boundary
value problem~(\ref{SDSE}) with zero boundary conditions at origin and at infinity.
The scattering amplitude is then defined by Eq.~(\ref{intrepr1}).

The theoretical results are supported by the numerical realization using the FEM-DVR technique.
{The theory is illustrated by application to both a one-channel as well as a two-channel problem which both include Coulomb interaction}.
Our formulation for the problem with the Coulomb interaction is theoretically as well as numerically
compared to the one expired by Rescigno {\it et al}~\cite{RBBC97}.
For multi channel scattering our analysis shows that the approach for the amplitude extraction
of Ref~\cite{RBBC97} gives the correct results for the cross sections in non-Coulomb channels and asymptotically correct results for
inelastic Coulomb channels.
{However, the representation for both the cross section and the scattering amplitudes which follows from the}
Rescigno {\it et al} formulation cannot be directly generalized for the Coulomb elastic channels.

For the practical implementation of our approach, we have introduced {a} cut-off radius for the short range potentials.
We have therefore estimated errors in the total cross section due to this cut off.
For the important class {of inverse} power potentials, we have found simple
estimations for the minimal cut-off radius which guarantees a desired accuracy.

\acknowledgments

The work of MVV, SLY and EY was supported by the Russian Foundation for Basic Research under the grant 08-02-01115-a.
SLY and EY are thankful to Stockholm University for the support of visit made possible under the bilateral agreement on cooperation between Stockholm University and St Petersburg State University.
This work was supported by grants from the Swedish National Research Council.

\end{document}